\documentclass[twocolumn,showpacs,preprintnumbers]{revtex4}

\def\be{\begin{equation}}
\def\ee{\end{equation}}
\def\beq{\begin{eqnarray}}
\def\eeq{\end{eqnarray}}
\def\n{\nonumber}
\def\bay{\begin{array}}
\def\eay{\end{array}}

\begin{document}

\preprint{CIRI/02-smw02}
\title{Spherical Gravitational Collapse and Accretion -\\ Exact General
Relativistic Description}

\author{Sanjay M. Wagh}
\affiliation{Central India Research
Institute, Post Box 606, Laxminagar, Nagpur 440 022, India\\
E-mail:ciri@vsnl.com}

\date{February 3, 2002}
\bigskip

\begin{abstract}

In this paper, we consider the problems of spherical gravitational
collapse and accretion using a spherically symmetric, spatially
homothetic spacetime, that is, as an exact solution \cite{cqg1} of
the field equations of general relativity. Properties of matter
like its equation of state determine whether the collapse becomes
unstoppable or not since the spacetime under consideration admits
any equation of state for matter in it. We can therefore describe
the formation of a semi-stable object here.  {\bf A black hole may
form in the unstoppable gravitational collapse and/or accretion
but only as an infinite red-shift surface that is, however, not a
null hyper-surface.} Therefore, spherical, astrophysical black
holes will always be of this type. This result will have important
implications for observational astrophysics and other
considerations in General Relativity based on the conception of
black hole as a null hyper-surface.
\\
\centerline{Submitted to: Physical Review D}
\end{abstract}

\maketitle

\newpage
\section{Introduction} \label{intro}
A {\em physically realistic\/} gravitational collapse problem
imagines matter, with regular initial data, collapsing under its
self-gravity. The resultant compression of matter causes pressure
to build-up in it. Further, matter compression generates heat and
radiation because of either the onset of thermonuclear fusion
reactions or other reasons. The radiation or heat then propagates
through the space. The collapsing matter could stabilize to some
size when its equation of state is such as to provide pressure
support against gravity. If self-gravity dominates, the collapse
continues to a spacetime singularity. The issue of Cosmic
Censorship Hypothesis \cite{penrose69} relates to whether the
singularity is visible to any observer or not, ie, whether it is
naked or not.

Irrespective of whether a black hole or a matter condensate forms
in the collapse, matter in the surroundings will accrete onto the
central object. The accreting matter may, initially, be dust in
the far away regions. However, it gets compressed as it moves
closer to the central object and pressure must build up in it. In
many such situations, heat and radiation partly escape the system
and partly fall onto the central object together with the
accreting matter.

Then, various stages of gravitational collapse can be
distinguished on the basis of the properties of matter such as its
equation of state that are different in different such stages.
Therefore, any complete description of the collapse and accretion
processes requires us to properly match spacetimes of various such
stages to produce the final spacetime description. Note that the
final spacetime will have to be a solution of the Einstein field
equations. (Note that the equation of state at extremely high
densities is not known.)

In any case, the final spacetime description of the gravitational
collapse and/or the accretion process must admit a changing
equation of state for collapsing/accreting matter. Further, such a
spacetime must also admit an energy or heat flux during late
collapse or accretion stages. To accomplish this process of
matching different such spacetimes is a herculean, if not
impossible, task.

Hence, another approach to this problem is essential. We could
then demand that a spacetime describing the collapse and/or the
process of accretion in its totality admits {\em any\/} equation
of state and appropriate energy-momentum fluxes. In other words,
the spacetime geometry should be obtainable from considerations
that do not involve the equation of state for the matter in the
spacetime. Furthermore, these considerations should result in a
spacetime admitting energy-momentum fluxes. We now turn to
precisely such considerations in General Relativity.

\subsection*{Spatially Homothetic Spacetimes}
The phenomenon of gravitation does not provide any length-scale or
mass-scale for spatial distribution of matter properties. Note
that the scale-independence of Newtonian gravity applies only to
space and not to time. This is one of the fundamental,
observational properties of gravitation. Therefore, General
Relativity as a theory of gravitation must not provide any
length-scale for matter distributions. This is essentially the
spatial scale-invariance property of the spacetime.

The scale-independence of gravity means that we can construct a
gravitating object of any size and of any mass. Further, matter
within such an object can be distributed in any desirable manner
since gravity does not provide for the spatial distribution of
matter within any gravitating object. (It is a separate question
as to whether every such object will be stable or not.)

There must therefore exist in General Relativity a spacetime that
allows matter density to be an arbitrary function of {\em each\/}
of the three spatial coordinates. We emphasize that such a
spacetime metric and all other metric forms that are reducible to
it under non-singular coordinate transformations are the only
solutions of the field equations of General Relativity that are
consistent with gravity not possessing a length-scale for matter
properties.

All other spacetimes that are {\em not\/} reducible to the
aforementioned spacetime by non-singular coordinate
transformations then {\em violate\/} this basic property of
gravity that it has no length-scale for matter properties. Hence,
such spacetimes must possess a specific length-scale or mass-scale
for matter properties. In short, not every solution of the
Einstein field equations respects the principle, namely, its
spatial scale-independence.

The field equations of General Relativity are based on Einstein's
equivalence principle which is, primarily, the principle of
equality of the {\em inertial\/} and {\it gravitational\/} masses.
It is the equivalence principle that leads to geometrization of
gravity and, hence, from a variational principle, to the field
equations of General Relativity. However, General Relativity does
not automatically incorporate other basic properties of gravity,
if any, for example that it does not specify any length-scale for
matter properties. Therefore, we need to {\em separately\/}
enforce the principle of ``no length-scale for matter properties''
on the solutions of the field equations of General Relativity.

In general, a homothetic Killing vector captures \cite{carrcoley}
the notion of the scale-invariance. The principle that gravity
does not specify any length-scale for matter properties then
requires the spacetime to admit, in general, three independent
spacelike Homothetic Killing vectors corresponding to the three
spatial dimensions. A spacetime that conforms to the spatial
scale-invariance, to be called a {\em spatially homothetic
spacetime}, is then required to admit, corresponding to each
spatial coordinate, an appropriate {\em spatial\/} homothetic
Killing vector ${\bf X}$ satisfying \be {\cal L}_{\bf X}
g_{ab}\;=\;2\,\Phi\,g_{ab} \ee where $\Phi$ is an arbitrary
constant. We then expect spatially homothetic spacetimes to
possess arbitrary spatial characteristics for matter. This is also
the broadest (Lie) sense of the scale-invariance of the spacetime
leading not only to the reduction of the Einstein field equations
as partial differential equations to ordinary differential
equations but leading also to their separation.

There is another importance of the spatially homothetic
spacetimes. We note that the field equations of General Relativity
were arrived at by demanding only that these reduce to the
Newton-Poisson equation in the weak gravity limit \cite{eg1913,
subtle}. But, the {\em field equations of any theory of gravity
should contain the entire weak gravity physics due to the
applicability of the laws of weak gravity to any form of matter
displaying any physical phenomena}. The field equations are
expected to be only the {\em formal\/} equality of the appropriate
tensor from the geometry and the energy-momentum tensor of matter.
Then, the field equations of General Relativity could have been
obtained by imposing the requirement that these reduce to the
single ``equation of the entire weak gravity physics''.

However, there is no ``single'' equation for the ``entire weak
gravity physics" since we include different physical effects in an
ad-hoc manner in the newtonian physics.

But, {\em there can be a ``single'' spacetime containing the
entire weak gravity physics}. Therefore, we need a principle to
identify such a solution of the field equations. In the weak field
limit, the spatial scale-invariance is the freedom of
specification of matter properties through three independent
functions of the three spatial coordinates, in general.

{\em The spatial scale-invariance is then the principle that could
help us identify spacetimes containing the entire weak gravity
physics.} Indeed. the spatial scale invariance identifies a single
such spacetime \cite{cch}. It has appropriate energy-momentum
fluxes, applicability to any form of matter and, hence, it
contains the entire weak gravity physics. The newtonian law of
gravitation then gets replaced by the single general relativistic
spacetime \cite{cch} that contains all of the weak gravity
physics. But, spatial scale-independence needs to be separately
imposed on the field equations to obtain it.

Spatially homothetic spacetimes do not yield a naked singularity
for initially non-singular, regular, spatial data for matter
fields \cite{cch}. Such spacetimes always produce a black hole in
the unstoppable gravitational collapse of spatially non-singular
matter of any spatial properties. Due to the freedom of the
specification of spatial properties of matter in it, a spatially
homothetic spacetime can describe the formation of some
gravitating object, may that be a black hole, and accretion onto
it in its entirety.

A general spatially homothetic spacetime, although known
\cite{cch}, is a complicated spacetime. The general problems of
the formation of a gravitating object and accretion onto it are,
therefore, highly involved problems to analyze in details. This is
understandable since, in general, a collapsing newtonian object of
arbitrary spatial characteristics will exhibit energy-momentum
fluxes along all the cartesian coordinate directions. We will
therefore have to deal with this complexity in all its generality.
It is instructive, however, to begin with the simpler situation of
spherical symmetry to gain insight into the physical nature of the
gravitational collapse problem.

To fix ideas, we therefore begin in this paper with the simplest
of such problems  - a spherically symmetric problem - that
considers the formation of a spherical gravitating object and
accretion of matter onto it. To this end, we first recall from
\cite{cqg1} the spherically symmetric, spatially homothetic
spacetime and its properties. It will be seen that the temporal
metric functions are determined by the properties of matter in it.

\section{Spacetime of Accreting, Non-Rotating, Spherical Object} \label{spacetime}
A general spherically symmetric spacetime admits a metric of the
form
\begin{widetext} \be ds^2 = -\,A^2(r,t)\,dt^2
\;+\;B^2(r,t)\,dr^2\;+\;C^2(r,t)\,\left[
\,d\theta^2\,+\,\sin^2{\theta}\,d\phi^2\,\right] \label{genssmet}
\ee \end{widetext} A spherically symmetric spacetime has only one
spatial scale associated with it - the radial distance scale. The
scale-independence of gravity then means that a spherically
symmetric spacetime allows {\em arbitrary\/} radial properties for
matter. The corresponding spatial homothetic Killing vector must
then possess only a radial component in appropriate coordinates.
Therefore, we impose \cite{cqg2} a {\em spatial\/}
homothetic Killing vector 
\be (\,0,\, f(r,t),\, 0,\, 0\,)\label{hkvss} \ee 
on the general spherically symmetric metric (\ref{genssmet}). This
{\it uniquely\/} determines the spherically symmetric metric to
that obtained in \cite{cqg1}, namely
\begin{widetext} \be ds^2=-\,y^2(r)\,dt^2+\gamma^2 \left(
y'\right)^2B^2(t)\,dr^2+y^2(r)\,Y^2(t)\,\left[
d\theta^2+\sin^2{\theta}d\phi^2 \right] \label{ssmetfinal}\ee
\end{widetext} with $f(r,t) = y/(\gamma y')$, a prime
indicating a derivative with respect to $r$ and $\gamma$ being a
constant. (We shall always absorb the temporal function in
$g_{tt}$ by suitable redefinition of the time coordinate.)

The coordinates are co-moving with the geometry. We can therefore
define \be D_t\;\equiv\;U^a\frac{\partial}{\partial x^a}
\;=\;\frac{1}{y}\frac{\partial}{\partial t} \ee where $U^a$ is the
four-velocity of the co-moving observer. Further, differentiation
along an outward radial unit vector orthogonal to $U^a$ is given
by \be D_r \;\equiv\;\frac{1}{\gamma (y') B}\,
\frac{\partial}{\partial r} \ee Then, the velocity of the fluid
with respect to the co-moving observer is

\be V_r \;=\;D_t\left(yY\right)\;=\;\dot{Y} \ee

\noindent The metric function $B(t)$ determines or gets determined
by the energy flux in the spacetime. These characteristics of the
temporal metric functions are important to the analysis of the
physics of the gravitational collapse implied by
(\ref{ssmetfinal}).

The Einstein tensor for (\ref{ssmetfinal}) has the following
components in the chosen coordinate frame
\begin{widetext} \beq G_{tt}&=& \frac{1}{Y^2}-\frac{1}{\gamma^2B^2} +
\frac{\dot{Y}^2}{Y^2} + 2\frac{\dot{B}\dot{Y}}{BY}
\\ G_{rr}&=&\gamma^2B^2
\left(\frac{y'}{y}\right)^2
\left[-\,2\frac{\ddot{Y}}{Y}-\frac{\dot{Y}^2}{Y}
+\frac{3}{\gamma^2B^2}- \frac{1}{Y^2}\right]
\\G_{\theta\theta}&=&-\,Y\,\ddot{Y}-Y^2\frac{\ddot{B}}{B}
- Y\,\frac{\dot{Y}\dot{B}}{B}+\frac{Y^2}{\gamma^2B^2}
\\G_{\phi\phi}&=& \sin^2{\theta}\,G_{\theta\theta} \\
G_{tr}&=&2\frac{\dot{B}y'}{By}  \eeq \end{widetext} where an
overhead dot denotes a time derivative.

Notice that the $t$--$r$ component of the Einstein tensor is
non-vanishing. Hence, the matter in the spacetime could be {\em
imperfect\/} or {\em anisotropic\/} indicating that the
energy-momentum tensor could be either of the following
\begin{widetext} \beq {}^{\rm
I}T_{ab}&=&(\,p\,+\,\rho\,)\,U_a\,U_b \;+\; p \,g_{ab}
\;+\;q_a\,U_b \;+\; q_b\,U_a \;-\;2\,\eta\,\sigma_{ab} \\
{}^{\rm A}T_{ab}&=&\rho \,U_a\,U_b \;+\; p_{||}\,n_a\,n_b \;+\;
p_{\bot}\,P_{ab} \eeq
\end{widetext} where $U^a$ is the matter four-velocity, $q^a$ is the
heat-flux four-vector relative to $U^a$, $\eta$ is the
shear-viscosity coefficient, $\sigma_{ab}$ is the shear tensor,
$n^a$ is a unit spacelike four-vector orthogonal to $U^a$,
$P_{ab}$ is the projection tensor onto the two-plane orthogonal to
$U^a$ and $n^a$, $p_{||}$ denotes pressure parallel to and
$p_{\bot}$ denotes pressure perpendicular to $n^a$. Also, $p$ is
the isotropic pressure and $\rho$ is the energy density.

For the observer co-moving with geometry with four-velocity \be
U^a = \frac{1}{y}\;{\delta^{a}}_t \ee the kinematical quantities
for the line element (\ref{ssmetfinal}) are given by \be {\dot
U}_a \;=\; {U_a}_{;\,b}U^b\;=\;\left(\,0,\, \frac{y'}{y},\, 0,\,
0\,\right) \label{acceleration}\ee \beq \Theta_M &=&
\frac{1}{y}\,\left(\,\frac{\dot{B}}{B} \;+\;
2\,\frac{\dot{Y}}{Y}\,\right) \label{expansion} \\ \sigma &\equiv&
\,{{\sigma}^{3}}_{3} = {{\sigma}^{2}}_{2} = -
\frac{1}{2}{{\sigma}^{1}}_{1}  \n \\
&=&\frac{1}{3\,y\,(2\eta)}\left(\,\frac{\dot{Y}}{Y} -
\frac{\dot{B}}{B}\,\right) \label{shearscalar} \eeq \smallskip

\noindent where $\dot{U}_a$ denotes the four-acceleration of
matter, $\Theta_M$ represents expansion of matter. Note that the
shear tensor is trace-free and $\sigma$ represents the
shear-scalar that is given by $\sqrt{6}\;\sigma$.

For $\Theta_M\,>\,0$, the spacetime under consideration is
expanding and, for $\Theta_M\,<\,0$, the spacetime is contracting.

Now, the Einstein field equations with imperfect matter yield for
(\ref{ssmetfinal})
\begin{widetext} \beq \rho &=&
\frac{1}{y^2}\left(\frac{\dot{Y}^2}{Y^2} + 2 \frac{\dot{B}}{B}
\frac{\dot{Y}}{Y} + \frac{1}{Y^2} - \frac{1}{\gamma^2 B^2}\right)
\label{sepdens} \\2\,\frac{\ddot{Y}}{Y} \;+\;\frac{\ddot{B}}{B}
&=&\frac{2}{\gamma^2B^2} \;-\; \frac{y^2}{2}\,\left(
\,\rho\,+\,3\,p \right) \label{isopressure}  \\
3\,(2\,\eta)\,\sigma &=&\frac{1}{y^2} \left(\,
\frac{\ddot{B}}{B}\;-\;\frac{\ddot{Y}}{Y}\;+\;\frac{\dot{B}\dot{Y}}{BY}
\;-\;\frac{\dot{Y}^2}{Y^2}\;+\;\frac{2}{\gamma^2B^2}\;-\;\frac{1}{Y^2}\,
\right) \label{ssshear} \\ q &=& - \frac{2 \dot{B}}{y^2 \gamma^2
y' B^3} \label{heatflux} \eeq
\end{widetext} where $q^a = (0, q, 0, 0)$ is the radial heat-flux
vector.

Clearly, the radial function $y(r)$ is not determined by the field
equations. Therefore, radial attributes of matter are {\em
arbitrary}, meaning, unspecified, for the metric
(\ref{ssmetfinal}). This is in the manner of concentric spheres
with each sphere allowed to possess any value of density, for
example. This is the maximal freedom compatible with the
assumption of spherical symmetry, we may note.

It also follows that the temporal metric functions $B(t)$ and
$Y(t)$ get determined by the properties of matter such as an
equation of state.

The point $r\,=\,0$ will possess a locally flat neighborhood when
${y'|}_{r\,\sim\,0}\;\approx\;1/\gamma$. This condition must be
imposed on any $y(r)$. Apart from this condition, the function
$y(r)$ is arbitrary. Other physical considerations, such as those
arising from the equation of heat transfer in the spacetime, could
constrain the function $y(r)$.

The density is, for $y'\,>\,0$, a decreasing function of $r$
corresponding to a region over-dense at its center. Therefore, for
our purposes here, we will assume that there is only one
over-dense region and, hence, $y'\,>\,0$ throughout the spacetime.
Then, we have that there is a ``single'' collapsing and accreting
spherically symmetric object.

The spatial or radial nature of the heat flux is determined
primarily by the sign of the quantity $-\,\dot{B}/y'$. The heat
flux is positive, that is, heat flows from lower values of $r$ to
higher values of $r$, when $y^{\prime}$ and $\dot{B}$ have
opposite signs. Then, with $y'\,>\,0$, we require $\dot{B}\,<\,0$
for the whole of the spacetime. Heat then flows from smaller
values of $r$ to larger values of $r$. That is to say, heat flows
from the central over-dense region to under-dense region
surrounding it. This is then the general gravitational model with
radially outward heat flux. A specific model is obtained for a
specific choice of the radial function $y(r)$ with above
conditions.

We remind the reader that additional conditions arising from the
considerations of the stability of the stellar object etc.\ will
constrain the radial metric function $y(r)$ in a manner similar to
those obtainable for the Newtonian model of a star.

In general, we may define the mass function by \be m(r,t)
\;=\;\frac{yY}{2}\,\left(\,1\,-\,\frac{Y^2}{\gamma^2B^2}\,+\,\dot{Y}^2\,\right)
\label{mass} \ee The field equations then imply \beq
\frac{\partial m}{\partial r} &=& 4\pi\,\rho\,y^2Y^3\,y' \\
\frac{\partial m}{\partial t} &=& -\,4\pi\,p\,y^3Y^2\,\dot{Y}
\label{accrate} \eeq For positive pressure $p$ in the spacetime
and for $\dot{Y}\,<\,0$, the mass accretes to the center and it is
increasing in time in a collapsing situation. We may also define
the luminosity of the central star as seen by a co-moving observer
at location $r$ by \be {\cal L}\;=\;4\,\pi\,y^2\,Y^2\,q\ee

\subsection*{Semi-stable, radiating object}
Recall that $\dot{Y}$ is the radial velocity of matter with
respect to the observer co-moving with the geometry. It can be
positive for out-flowing matter, negative for in-flowing matter
and zero for stable matter. Also, the mass accretion rate
(\ref{accrate}) will vanish for $\dot{Y}\;=\;0$. Then, the
temporal dependence of the mass function in (\ref{mass})
corresponding  to $\dot{Y}\;=\;0$ is purely due to the conversion
of mass to radiation or heat. In this case, we also obtain \be
(2\eta)\,6\,\sigma\;=\;\gamma^2 B^2\,y\,y'\,q \ee for
$\dot{Y}\,=\,0$.

This is the ``semi-stable'' spherically symmetric, radiating
object. The expressions for its density etc.\ follow from
(\ref{sepdens}) - (\ref{heatflux}). In particular, we can write
\beq \rho
&=&\frac{1}{y^2}\left(\,1\;-\;\frac{(y')y^2B}{2|\dot{B}|}\,q\,\right)
\\\frac{\ddot{B}}{B}&=&\frac{2}{\gamma^2B^2} \,-\,\frac{y^2}{2}
\left(\rho+3\,p\right)\eeq by noting that the positivity of shear
and  heat flux, both, requires $\dot{B}\,<\,0$ for $y'\,>\,0$.

Note that the stability of such an object is not for all of the
co-moving time. The properties of matter determine whether the
object remains stable in this manner or not. Considerations such
as those leading to the Chandrasekhar or the Oppenheimer-Volkov
limits \cite{stars} are then possible. However, these require more
details of the matter properties than are considered here.

\subsection*{No collapse without heat generation}
We may note that the heat generation at some stage during the
gravitational collapse is expected on the basis of very general
physical considerations of thermodynamic origin. Therefore, we
must not obtain the situation of gravitational collapse without
heat generation when we use the metric (\ref{ssmetfinal}).

This is easily seen by taking $\dot{B}\,=\,0$ so that the heat
flux vanishes in the spacetime for all co-moving time since there
is no heat generation. But, (\ref{shearscalar}) implies that
$\dot{Y}\,>\,0$ for the shear-scalar to be positive. Then,
(\ref{expansion}) implies that matter in the spacetime of
(\ref{ssmetfinal}) is expanding and not contracting.

Consequently, we do not obtain the situation of gravitational
collapse in the absence of heat generation in the spacetime of
(\ref{ssmetfinal}).

\subsection*{Shell-crossing and Shell-focussing singularities}
The singularities at locations for which $y'=0$ are of
shell-crossing type. These are however weak singularities since
the curvature invariants do not blow up at locations for which
$y'=0$  \cite{cqg1}. Such locations have physical meaning in terms
of the heat flow caustics.

The genuine spacetime singularities of the strong curvature,
shell-focussing type exist when either $y(r)\,=\,0$ for some $r$
or when the temporal functions vanish for some $t$. The
possibility of $y(r)\,=\,0$ for some $r$, however, means that we
already have a spacetime singularity at that $r$. Therefore, we
have, for spherically symmetric, spatially homothetic spacetimes,
that $y(r)\,\neq\,0$ at all $r$ for the non-singular initial data
for matter fields.

(In general, for spatially homothetic spacetimes, the
non-singularity of initial data for matter fields will be seen
\cite{cch} to require the non-vanishing of the corresponding
arbitrary functions of the spatial coordinates.)

Further, we note that the ``physical'' radial distance
corresponding to the ``coordinate'' radial distance $\delta r$ is
\be \ell\;=\;\gamma (y') B \delta r \ee Then, a collapsing shell
of matter forms the spacetime singularity in the present spacetime
when $B(t)\,=\,0$ is reached for it at some $t\,=\,t_s$. The
temporal function $Y(t)$ determines the shear and the expansion in
the spacetime together with the temporal function $B(t)$.

\subsection*{Light Trapping Surface} \label{light-trap}
A radially outgoing null vector of (\ref{ssmetfinal}) is \be
\ell^a\partial_a\;=\frac{1}{y}\frac{\partial}{\partial t}
\;+\;\frac{1}{\gamma y' B}\frac{\partial}{\partial r}\,=\,D_t +
D_r \label{rnull} \ee Light gets trapped inside a particular
radial coordinate $r$ when the expansion of the above principle
null vector vanishes. The formation of the outermost {\em
light-trapping surface\/} (LTS) during any unstoppable collapse is
then obtained by setting the expansion of (\ref{rnull}) to zero.

The zero-expansion of (\ref{rnull}) yields a condition only on the
temporal metric functions as \be
y\,\Theta_M\,\equiv\,\frac{\dot{B}}{B}\,+\,2\,\frac{\dot{Y}}{Y}\;
=\; -\frac{3}{\gamma B} \label{ltscon} \ee When this condition is
reached during the gravitational collapse light and, with it,
matter trapping occurs. Note however that we are dealing here with
the picture of the co-moving observer. For the co-moving observer
$\dot{Y}$ is the radial velocity of matter and $\dot{B}$
determines the heat flux.

\subsection*{No null hyper-surface or horizon}
Note that (\ref{ltscon}) is not the condition for the formation of
a null hyper-surface or the horizon in the spacetime. The
formation of horizon or a null hyper-surface requires the normal
of a hyper-surface to become null.

The existence of a spherical null hyper-surface requires that the
norm of its normal vanishes at some $r$. If $n_a\,=\,(0, 1, 0, 0)$
is the normal to $r\,=\,$ constant hyper-surface, then we have
$n^an_a\,=\,1/(\gamma^2 (y')^2B^2)$ which, obviously, cannot
vanish except for $y' \rightarrow \infty$ for some $r$ - an
evidently degenerate-metric situation.

Moreover, we may, in terms of the mass function of (\ref{mass}),
write \be
g_{rr}\;=\;\frac{Y^2\,(y')^2}{1\,+\,V_r^2\,-\,\left(2m/yY\right)}
\ee Then, $g_{rr}\rightarrow \infty$ implies $Y(t)\,\rightarrow
0$. Importantly, therefore, there cannot form a spherical null
hyper-surface at any radial location in (\ref{ssmetfinal}).

Also, the coordinate speed of light in the spacetime of
(\ref{ssmetfinal}) is \be \frac{dr}{dt} \;=\;\pm\,\frac{y}{\gamma
(y') B} \label{cspeed} \ee This speed cannot vanish for any $r$
except in the case of an initially singular density distribution
at $r$ which we do not consider to be any serious, astrophysically
meaningful, initial condition here.

Therefore, a null hyper-surface or horizon does not form in
(\ref{ssmetfinal}) for any non-singular initial data for matter
fields.
\subsection*{But, ``Black Hole'' forms}
However, the light-trapping properties of gravity exist in the
sense that gravity becomes strong enough to trap light in a
spacetime region when the condition (\ref{ltscon}) is satisfied.
(See also next section.)

We recall again that $\dot{Y}$ is the velocity of the fluid
relative to the co-moving observer and that the temporal function
$B(t)$ determines or is determined by the heat generation in the
spacetime. The properties of matter then determine the temporal
metric functions in the spacetime of (\ref{ssmetfinal}).
Therefore, depending on the properties of matter in the spacetime,
trapping of light and matter occurs for (\ref{ssmetfinal}).

The above is understandable as follows. A spherical star begins to
collapse and the velocity of its matter increases as its collapse
continues in the frame of the co-moving observer. It is only at
some ``instant'' of the co-moving time that the curvature becomes
strong enough to trap light and matter. The condition
(\ref{ltscon}) determines this instant of the co-moving time.

This is the formation of a black hole at an instant of the
co-moving time at which the light and, with it, matter get trapped
in a strong gravitational field. This is the conception of a black
hole that applies here and not that of a null hyper-surface, we
may then note.

\section{Spherical collapse, accretion and black hole formation} \label{ssaccrete}

In the usual analysis of accretion onto a gravitating object, we
generally consider a ``central'' object (that has already formed)
and the surrounding matter (which is accreting onto it). However,
this picture gets replaced by the problem of the formation of the
central object and continued pile up of matter to the central
region when we use the spatially homothetic spacetimes. This is
primarily because these spacetimes have no spatial length scale
and hence contain matter everywhere.

With the above feature of the spatially homothetic spacetimes in
mind, we now turn to the problem of the continued accretion of
matter onto a central object in the spherical collapse.

An observer co-moving with the geometry evaluates ``various''
quantities and interprets the observations of the star ``as an
asymptotic observer'' for large $r$. On the other hand, an
observer in the rest frame of the accreting matter is also
important to the physical analysis of the problem under
consideration. We therefore analyze the problem at hand by
considering the observer in the rest frame of matter.

\subsection*{In the rest frame of matter}
The four-velocity of the matter fluid with respect to the
co-moving observer is: \be
U^a\;=\;\left(\,U^t,\,U^r,\,0,\,0\,\right) \ee Defining then the
radial velocity of matter with respect to the co-moving observer
as \be V_r\,\equiv\,U^r/U^t
\ee we then obtain from the metric (\ref{ssmetfinal}):
\beq U^a &=& \frac{1}{y\,\sqrt{\Delta}}\,\left(\,1,\,V_r,\,0,\,0\,\right) \\
\Delta &=&\;1\;-\;\gamma^2\,\left(
\frac{y'}{y}\right)^2\,B^2\,V_r^2 \label{Delta} \eeq

Now, if $d\tau_{\scriptscriptstyle CM}$ is a small time duration
for the co-moving observer and if $d\tau_{\scriptscriptstyle RF}$
is the corresponding time duration for the observer in the rest
frame of matter, then we have \be d\tau_{\scriptscriptstyle
CM}\;=\;\frac{d\tau_{\scriptscriptstyle RF}}{\sqrt{\Delta}}
\label{rshift} \ee Therefore, the co-moving observer waits for an
infinite period of its time to receive a signal from the
rest-frame observer when $\Delta\,=\,0$. Equation (\ref{rshift})
is also the red-shift formula. Clearly, therefore, $\Delta\;=\;0$
is the {\em infinite red-shift surface}.

But, the above infinite red-shift surface is {\em not\/} a one-way
membrane or a null hyper-surface since $U^a\,U_a\,=\,-1$ always.
{\em This is an important point of distinction for the spatially
homothetic spacetime of (\ref{ssmetfinal}) from the spacetimes
violating the spatial scale-invariance of gravity eg, the
Schwarzschild spacetime for which a null hyper-surface exists and
the infinite red-shift surface is also a null hyper-surface.}

Of course, the infinite red-shift surface separates the spacetime
of (\ref{ssmetfinal}) into two regions - one that can communicate
to the far away zone and the black hole region that cannot. The
inside and outside of the infinite red-shift surface are then
causally disconnected regions of the spacetime of
(\ref{ssmetfinal}).

We have then the following possibilities \beq
(\Delta\,>\,0)\qquad |\gamma\,(y')\,B\,V_r| \;<\;y \label{bh0} \\
(\Delta\,=\,0)\qquad |\gamma\,(y')\,B\,V_r| \;=\;y \label{bh1} \\
(\Delta\,<\,0)\qquad |\gamma\,(y')\,B\,V_r| \;>\;y \label{bh2}
\eeq Matter with an initial density distribution determined by
$y(r)$ begins to collapse under the condition (\ref{bh0}) with
initial velocity $V_{r, ini}$ and initial heat flux, determined by
$B$, being small. The in-fall velocity of matter and heat flux
grow as matter collapse progresses. Matter properties decide
whether the collapse becomes unstoppable or not. Then, regions
satisfying (\ref{bh1}) and (\ref{bh2}) form when condition
(\ref{ltscon}) is reached.

In the case of accretion process, matter with inwardly directed
radial velocity $V_r\,=\,-\,|\dot{Y}|$ and $\Delta \,>\,0$
accretes onto the central object, may that be any, under suitable
conditions such as the radiation pressure not reversing its radial
velocity etc. (This is how the properties of matter play an
important role in the collapse or accretion processes.) The radial
in-fall velocity of matter increases with the time of the
co-moving observer in any unstoppable collapse. The condition
$\Delta \,=\,0$ is then reached for some part of matter in the
spacetime of (\ref{ssmetfinal}) when the temporal metric function
$B(t)$ satisfies (\ref{ltscon}).

To continue with the analysis of the light trapping surface, we
note that conditions (\ref{bh0}) - (\ref{bh2}) possess an
immediate interpretation. Noticing that the ``physical'' radial
distance is $\gamma (y')B$, these conditions imply that the
physical distance covered per unit co-moving time with velocity
$V_r$ is less than, equal to or greater than the distance measured
by $y$. That is, if we use $y$ as the radial coordinate (with
$y(r=0)\,=\,y_c\,\neq\,0$), the conditions (\ref{bh0}) -
(\ref{bh2}) have the above interpretation. In particular, at the
outermost Light Trapping Surface, we expect the in-fall velocity
$V_r\,=\,1$ and, hence, the condition (\ref{bh1}) implies that the
``coordinate'' distance $y$ equal the light-travel time from this
surface to the singularity at $B(t_s)\,=\,0$ at this surface. We
note that this is a very natural interpretation of the condition
(\ref{bh1}). (See, also, (\ref{cspeed}).)

Then, matter shells, in their rest frame, cross the Light Trapping
Surface in successions when the shell-labelling radial coordinate,
$y$, corresponds to the light-crossing time between the LTS and
the singularity at $B(t_s)\,=\,0$. Therefore, matter within the
region $y \,\leq\, \gamma (y') B$ gets trapped inside the outmost
light-trapping surface when the collapse advances to satisfy the
condition (\ref{ltscon}) on the temporal metric function $B$.

\subsection*{Initial conditions}
We note that the initial conditions for the spacetime of
(\ref{ssmetfinal}) consist of conditions at the ``initial
co-moving time'' and not for large radial distances. Primarily,
the `initial' temporal functions are to be chosen on the basis of
the choice of the radial function $y(r)$. If the `initial' radial
density distribution is such as to result to immediate heat
generation then the temporal metric function $B$ is not initially
constant. So, is the case with `initial' value of shear in the
spacetime. Therefore, the nature of `initial' temporal and spatial
metric functions is to be decided on the basis of the
`astrophysical' nature of the problem under consideration.

When we consider the gravitational collapse of matter from some
initial density distribution corresponding to a ``small''
over-density, the initial heat or radiation flow in the spacetime
could be very small. Then, in the absence of any ionizing
radiation, matter in the spacetime could be expected to be
``dusty'' ie, approximately pressureless with negligible heat flow
at initial co-moving time.

Moreover, matter may, at initial co-moving time, be assumed to be
non-relativistic throughout the spacetime and, hence, we have \be
V_r\, \equiv\, V_{ini}\;<<\;1 \qquad {\rm(early\,time)} \ee We may
also expect negligible radiation in the spacetime at initial
co-moving time, indicating that $B\,\approx$ constant initially.
Then, initially, \be \Delta \approx 1 - \gamma^2\left(
\frac{y'}{y}\right)^2B^2V_{ini}^2\,\approx\,1\ \ee

However, heat flow eventually grows. In this connection, we
emphasize that a star forms at some suitable co-moving time and
emits radiation that flows through the medium surrounding it. For
large distances from the star, the radiation flux is weak and,
hence, non-ionizing. It is then understandable that the spacetime
of (\ref{ssmetfinal}) does not admit a collapsing dust solution
without heat flux since radiation must flow in the spacetime at
large $r$.

It may be noted that the large $r$ condition of matter is to be
inferred on the basis of only the astrophysical expectation that
matter will be non-relativistic and dusty far away from a source
of intense radiation. This is the sort of situation that is
commonly observed with accreting objects.

We emphasize here that this expectation is, automatically, borne
out by the spacetime of (\ref{ssmetfinal}) on the basis of the
radial dependence of various physical quantities. The condition of
matter for large $r$ is, however, not essential to the physics of
the gravitational collapse beyond specifying the properties of
matter in the far-away zone. Then, close to the central object the
equation of state of matter is expected to be different. This is
precisely the situation with the spacetime of (\ref{ssmetfinal}).

Further, describing the fluid by its local thermodynamical
properties \cite{maartens}, the energy conservation principle
implies the first law of thermodynamics in the form \be
I_{,\,a}U^a\;=\;-\,{\cal
C}\,-\,p\,\left(\frac{1}{n}\right)_{,\,a}U^a \ee where we have
used $\rho\,=\,n\,(1+I)$ with $I$ as the specific internal energy,
${\cal C}(T, n)$ as the rate of decrease of internal energy per
unit amount of matter and $n$ as the number density of particles
of matter. The function ${\cal C}$ is related to the heat flux
$q$.

Note also that, in general, we can write \be
V_r\;=\;\frac{yY}{2}\,\left[\,(2\eta)\,3\,\sigma\,-\,\gamma^2B^2(yy')\,q
\,\right] \ee Some of the standard analysis \cite{book1} of the
accretion process can then be followed for the spacetime of
(\ref{ssmetfinal}) by considering appropriate initial conditions.

However, our purpose here is limited only to showing that a black
hole forms in the gravitational collapse only as an infinite
red-shift surface that is not a null hyper-surface. Therefore, we
do not consider here the details of this analysis which will,
however, be the subject of our future works. But, we have
presented here all the details which are essential to consider the
problem of spherical accretion in its totality.

\section{Concluding remarks}
In \cite{nonaked} we showed that naked singularities do not form
in spherically symmetric, spatially homothetic spacetimes for
non-singular, initial data for matter fields. In
\cite{sscollapse}, we considered the shear-free gravitational
collapse and its implications for the Cosmic Censorship
Hypothesis. In \cite{cqg1}, we showed that the Cosmic Censorship
Hypothesis is equivalent to the statement that gravity has no
length-scale for matter properties. Further to all of the above,
we showed in this paper that the spatially homothetic, spherically
symmetric spacetime of (\ref{ssmetfinal}) leads to a very general
description of the formation of a spherical object in General
Relativity. In particular, we showed that the spacetime of
(\ref{ssmetfinal}) admits a black hole only as an infinite
red-shift surface and not as a null hyper-surface.

Further, considering that the spacetime of (\ref{ssmetfinal})
admits any equation of state for matter in the spacetime and that
it is the only spherically symmetric spacetime satisfying the
spatial scale-invariance of gravity, we emphasize that black holes
obtained in spherical gravitational collapse in Nature must
necessarily be of the type considered here.

The description of spherical gravitational collapse, presented in
this paper, is also applicable to the problem of spherical
accretion onto a central gravitating object. We have displayed all
the equations necessary for any detailed analysis of this
important astrophysical problem that will be dealt with in future
subsequent works.

As a final remark, we note that many features of spherical
gravitational collapse and accretion processes considered here
will be obtainable for general spatially homothetic spacetime
\cite{cch}. In particular, the initial conditions will be in terms
of the temporal metric functions and the spatial density
distribution. The light trapping property of gravity in terms of
the formation of the light trapping surface will also correspond
to some ``instant'' of the co-moving time. A general spatially
homothetic spacetime will not admit a null hyper-surface but a
black hole can form in it only as an infinite red-shift surface.

The nature of the black hole that we obtain for spatially
homothetic spacetimes is very similar to that obtainable
\cite{laplace} in the newtonian theory of gravity except that our
considerations here are fully general relativistic. The black hole
of (\ref{ssmetfinal}) is only an infinite red-shift surface and
not a null hyper-surface. In general, the black hole of the
general spatially homothetic spacetime \cite{cch} can be expected
to be only an infinite red-shift surface. It is then to be noted
that these results will have important implications for
observational astrophysics and for other considerations in General
Relativity \cite{bhthermo1} - \cite{bhthermo4} that have been
based on the conception of a black hole as a null hyper-surface.

\end{document}